\documentclass{article}

\usepackage[nonatbib,preprint]{neurips_2020}
\usepackage{graphicx,subfig}
\usepackage[utf8]{inputenc} 
\usepackage[T1]{fontenc}    
\usepackage{hyperref}       
\usepackage{url}            
\usepackage{booktabs}       
\usepackage{amsfonts}       
\usepackage{nicefrac}       
\usepackage{microtype}      
\usepackage{amsmath,amsxtra,amsfonts,amscd,amssymb,bm,amsthm,svg}
\usepackage{bbm}
\usepackage{enumitem}
\title{Interpretable Neural Networks for Panel Data Analysis in Economics
	\thanks{Yang and Zheng contributed equally to this paper. We thank Bin Dong, Guanhua Huang, Dake Li, Chris Sims, Ranran Wang, Lingzhou Xue, Linfeng Zhang, Guang Zeng and audience in the BIBDR Economics and Big Data Workshop for helpful comments.}
}
\usepackage{multirow}
\usepackage{wrapfig}
\setlist[enumerate]{nosep}
\author{
	Yucheng Yang\\
	Princeton University\\
	\texttt{yuchengy@princeton.edu} \\
	\And
	Zhong Zheng \\
	Penn State University \\
	\texttt{zvz5337@psu.edu} \\
	\And
	Weinan E \\
	Princeton University \\
	\texttt{weinan@princeton.edu} \\
}

\newtheorem{prop}{\sc {Proposition}}[section]

\DeclareMathOperator{\E}{\mathbb{E}}

\begin{document}

\maketitle
  \vspace{-0.8cm}
\begin{center}
       November 2020
 \vspace{0.3cm}
\end{center}

\begin{abstract}
The lack of interpretability and transparency are preventing economists from using advanced tools like neural networks in their empirical research. In this paper, we propose a class of interpretable neural network models that can achieve both high prediction accuracy and interpretability. The model can be written as a simple function of a regularized number of interpretable features, which are outcomes of interpretable functions encoded in the neural network. Researchers can design different forms of interpretable functions based on the nature of their tasks. In particular, we encode a class of interpretable functions named \textit{persistent change filters} in the neural network to study time series cross-sectional data. We apply the model to predicting individual's monthly employment status using high-dimensional administrative data. We achieve an accuracy of 94.5\% in the test set, which is comparable to the best performed conventional machine learning methods. Furthermore, the interpretability of the model allows us to understand the mechanism that underlies the prediction: an individual's employment status is closely related to whether she pays different types of insurances. Our work is a useful step towards overcoming the ``black box'' problem of neural networks, and provide a new tool for economists to study administrative and proprietary big data.
\end{abstract}

\section{Introduction}
Traditionally, economists have relied on interpretable models like linear or logistic regressions, which provide clear insights on the causal or statistical relationships in small datasets \cite{angrist2008mostly, wooldridge2016introductory}. Recently, the use of administrative and proprietary big data has led to lots of exciting work in empirical economics \cite{einav2014economics,chetty2014land,mian2015house, saez2016wealth}. Though such datasets enjoy the richness of variables and large sample size, advanced tools like neural networks have not been adopted widely in their analysis as expected. Methodologically, there have been some attempts to bring machine learning tools to economic analysis. However, most of those successful applications are limited to relatively simple models like Lasso, ridge regressions or decision trees \cite{mullainathan2017machine,chernozhukov2018double,kleinberg2018human}. Economists are still nervous about using more advanced tools like neural networks, since they mostly deliver outcomes from a complicated black box without transparency or interpretability \cite{athey2018impact}, even though they enjoy better capability to fit data well than traditional econometric models.\par
\begin{figure}[h!]
\begin{center}
  \subfloat[Conventional Neural Networks]{\includegraphics[width=0.48\textwidth]{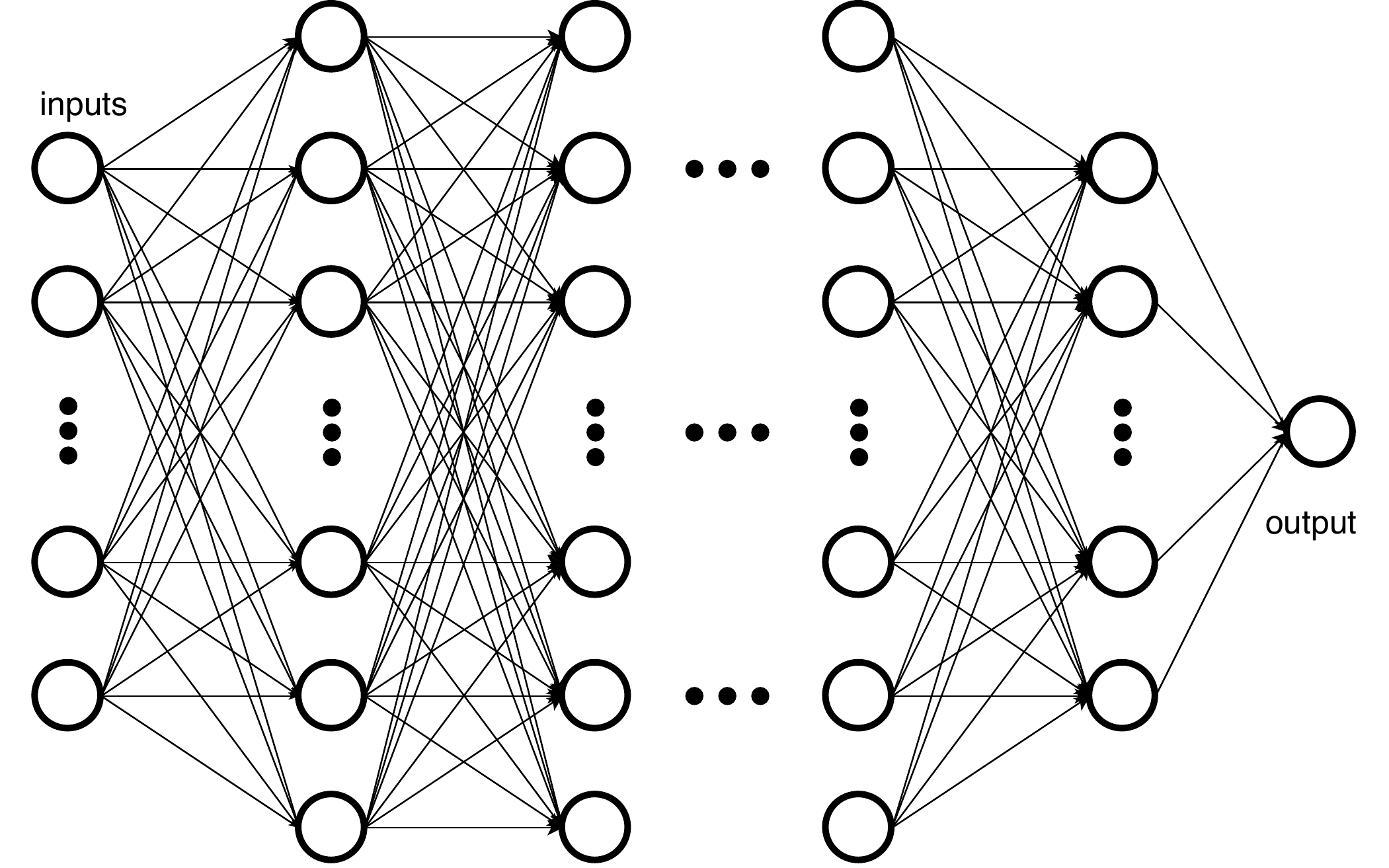}\label{fig:f1}}
  \hfill
  \subfloat[Interpretable Neural Networks]{\includegraphics[width=0.48\textwidth]{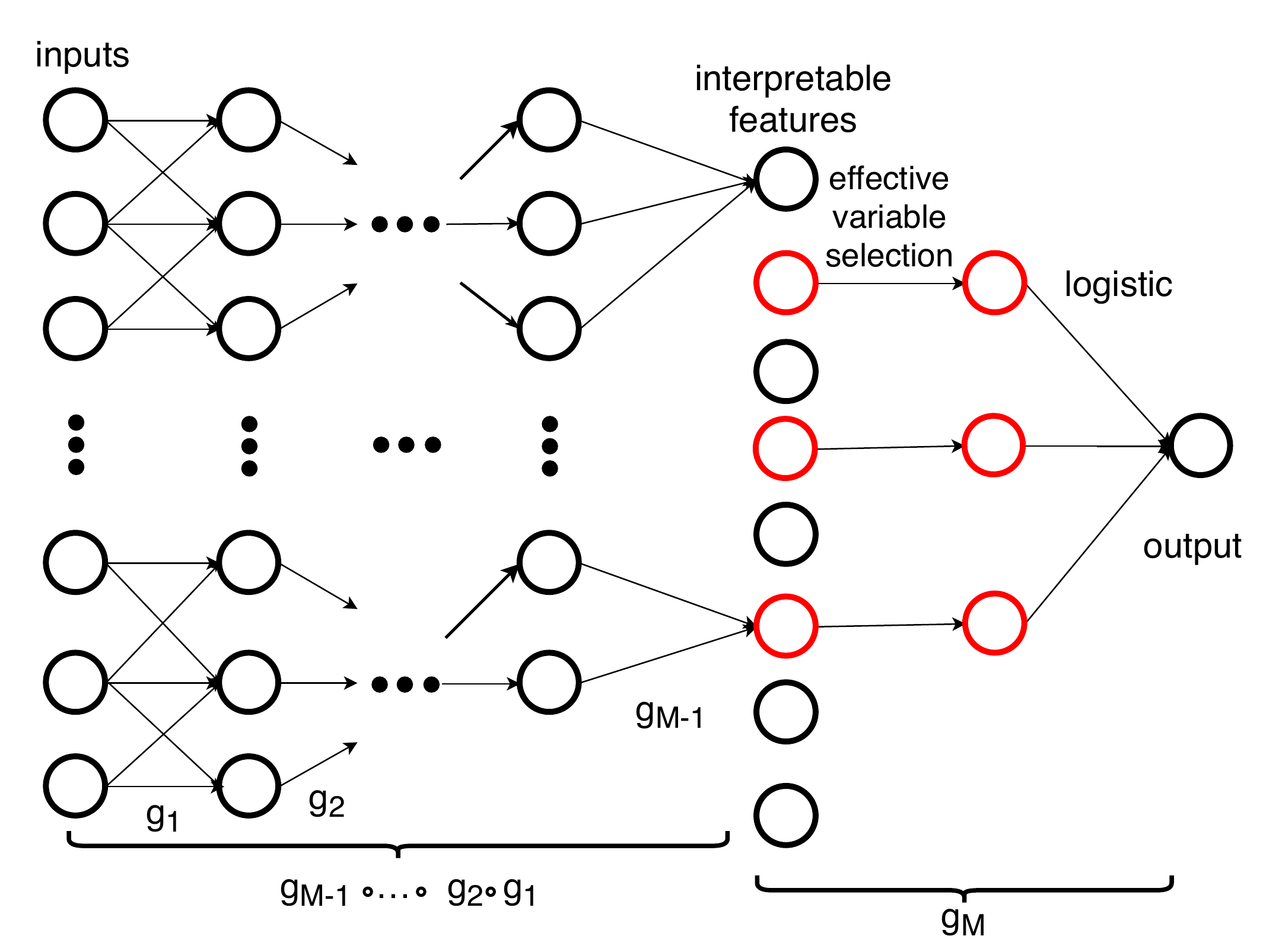}\label{fig:f2}}
  \caption{{\small Comparison of Conventional Neural Networks (left) and Interpretable Neural Networks (right).
  }}
  \label{fig:compare0}
  \end{center}
\end{figure}

To address this dilemma, we propose a new class of interpretable neural network models. Our model allows us to take advantage of  the high accuracy of neural networks, while remaining interpretable like linear or logistic regressions.
To this end, we design a modified version of neural networks: all layers except the last one encode various interpretable functional forms with unknown parameters, while the final layer is a logistic function that only takes a regularized number of neurons as inputs. As is shown in Figure \ref{fig:compare0}, our model can essentially be written as a simple logistic function of a limited number of interpretable features (red circles), which makes it more interpretable than conventional neural networks. Under this general formulation, researchers can design different interpretable functional forms based on their own tasks. In this paper, we design a class of \textit{persistent change filters} as part of the network, which turns out particularly helpful in time series cross-sectional data analysis\footnote{Time series cross-sectional data are ``data with a cross section of units with repeated observations on them over time''. It is also called panel data. In this paper, we will use these two names interchangeably.}. \\
\vspace{-1cm}

The interpretable neural network model we propose satisfies all the three desiderata for interpretable machine learning models \cite{murdoch2019definitions}. First, it has high ``predictive accuracy'' due to the capability of its multi-layer architecture to fit data well. Second, it has high ``descriptive accuracy'' since the model structure is designed to be interpretable. Third, it meets the ``relevancy'' requirement. Reseachers can design different interpretable functional forms in different tasks, so that the model estimates can deliver relevant knowledge. Beyond these three desiderata, our model estimates are robust to data missing problems, which is crucial if we expect to get reliable interpretations of the model.\par


As an application, we use the interpretable neural network models with \textit{persistent change filters} to predict individual's monthly employment status using high-dimensional administrative data in China. We achieve an accuracy of 94.5\% on the test set, which is comparable to the best-performing conventional machine learning methods. In addition, we clearly understand how the model predicts an individual's employment status with her payment records of different types of insurances. Both the accuracy and the interpretability are robust subject to data missing problems.\par
This paper contributes to the literature with the following distinct features. First, we expand the machine learning toolbox for economists \cite{mullainathan2017machine,chernozhukov2018double,kleinberg2018human,wager2018estimation} by introducing a modified neural network model that achieves high accuracy without sacrificing interpretability and transparency. Second, we contribute to the large literature on interpretable machine learning. Previous work in computer vision and natural language processing \cite{doshi2017towards, zhang2018interpretable, lei2017interpretable} mostly focus on interpreting how hidden units of neural networks represent local features of images and texts. In this paper, we propose an interpretable model to study panel data\footnote{Our paper is also distinct from recent literature that study causal inference in panel data with machine learning tools \cite{athey2018matrix,abadie2019using}. Our focus is to develop an interpretable model that achieve high predictive accuracy on the data.}, to help people understand the mechanisms behind the collective human behavior. In the terminology of the overview paper on interpretable machine learning \cite{murdoch2019definitions}, our model delivers ``model-based interpretability'', which is different from work that delivers ``post hoc interpretability'' by approximating outcomes from complicated models with simple functions. Last but not least, the application of our model to predicting individual employment status with China's administrative data also contributes to the large literature on China's unemployment study in the absence of reliable official statistics \cite{giles2005china,feng2017long}. Our work stands out in this literature as a novel application of machine learning methods on administrative big data, rather than traditional national account approach on household survey data.\par
The remaining of the paper is organized  as follows. In Section \ref{sec:model}, we present the architecture of the interpretable neural network model. In particular, we encode a class of interpretable functions named \textit{persistent change filters} in the model to study panel data. In Section \ref{sec:application} we exhibit the virtues of the model by applying it to predict individual's employment status in a large administrative dataset. Finally we conclude with discussions on the future work.

\section{Interpretable Neural Networks for Panel Data}\label{sec:model}
\subsection{Interpretable Neural Networks: General Formulation}\label{sec:general}
We first lay down the general formulation for the interpretable neural network model. For observation $i$, we write the model as $y_{i}= g(\boldsymbol{x}_{i},\theta)$, where $g$ shares the multi-layer structure as conventional neural networks $g = g^{(M)} \circ g^{(M-1)} \circ \dots \circ g^{(1)}$. \par
For the first $M-1$ layers, $g^{(M-1)} \circ \dots \circ g^{(1)}$ are multi-dimensional differentiable interpretable functions with unknown parameters. Here the functional forms should be designed based on the nature of tasks, since interpretable functions in computer vision may look very different from those in economics. These functions map the original variables into high-dimensional interpretable features. The last layer $g^{(M)}$ is a logistic function with these interpretable features as inputs. To make the final model more interpretable, we require the final layer to only take a small number of interpretable features as inputs. This is achieved by adding a Lasso \cite{tibshirani1996regression} or group Lasso \cite{yuan2006model} type of penalty on the parameters of the final layer to the loss function:
\begin{equation}\label{eqn:loss}
    L(\theta,\phi)=\E_{X,Y} \text{Loss}(g(x,\theta),y)+\lambda \mbox{penalty}(g^{(M)})
\end{equation}
Here $\lambda$ is a hyperparameter. The full architecture of interpretable neural network $g$ is in the right panel of Figure \ref{fig:compare0}. Since all components are differentiable in loss function (\ref{eqn:loss}), the model can be trained like a conventional neural network with gradient descent algorithms like Adam \cite{kingma2014adam}.
\subsection{Interpretable Neural Networks for Panel Data}\label{sec:math}
Now we design an interpretable functional form for panel data analysis. Consider a model $y_{it}= g(\mathbf{x}_{it},\theta) = g^{(M)} \circ g^{(M-1)} \dots \circ g^{(1)}$, where $y_{it}$ is the outcome for unit $i$ at time $t$ and $\mathbf{x}_{it}$ are high dimensional inputs. We choose $M = 4$ and design functional forms for $\{g^{(m)}\}, m = 1,2,3$ as below.
\begin{enumerate}[leftmargin=*]
\item $g^{(1)}$: Decision-tree-like splitting $\text{Sp}(x) = \mbox{Sigmoid}(c x+d)$.
For $x \in \mathbb{R}^1$, a sigmoid function can function as a decision-tree-like splitting operator like a differentiable indicator function \cite{yang2018deep}. 
For example, if $c$ is close to 1 and $d$ is a large negative number, the output can be close to 0 and 1 to discriminate whether $x$ is smaller or larger than a threshold. Here $c$ and $d$ are unknown parameters.
\item $g^{(2)}$: Dimension reduction $\text{Re}(\mathbf{x})$.  
For $\mathbf{x} \in \mathbb{R}^m$, define $\text{Re}(\mathbf{x}): \mathbb{R}^m \rightarrow \mathbb{R}^n (n \ll m)$ as a certain dimension reduction function with unknown parameters. For example, $\text{Re}(\mathbf{x})$ could be linear combination with sparse inputs, and the linear coefficients are unknown parameters.
\item $g^{(3)}$: \textit{persistent change filter} $\mathcal{D}(\mathbf{x})$. Now we introduce a class of interpretable functions named \textit{persistent change filter}, which turns out helpful in panel data analysis. $\mathcal{D}(\mathbf{x})$ also include unknown parameters.
\end{enumerate}
\subsection{A Class of Interpretable Functions for Time Series Data: \textit{Persistent Change Filter}}\label{sec:dist}
For a particular time series for some individual unit $i$ (the $i$ subscript is omitted here for simplicity), $x_\tau \in [0,1], \tau =1,2...t$,  we define the \textit{persistent change filter} $z_t = \mathcal{D}(\{x_\tau\}_{\tau =1}^t)=p_t-q_t$, where:
$$p_1=x_1,p_{\tau+1}= [1 + k(x_{\tau+1}-1)]p_{\tau} + x_{\tau+1}$$
$$q_1=1-x_1,q_{\tau+1}=[1 + k(-x_{\tau+1})]q_{\tau} + (1-x_{\tau+1})$$
Here $k\in[0,1]$ is an unknown parameter. Then we can map the original time series $\{x_t\}, t = 1, 2, ..., T$ to a \textit{persistent change filter} time series $\mathcal{D}(\{x_\tau\}_{\tau = 1}^t), t = 1, 2, ..., T$. The definition of the \textit{persistent change filter} $\mathcal{D}$ is motivated by identifying a persistent change in time series data. We illustrate this idea first with two examples and formalize with a proposition. 
\begin{figure}[h!]
\vspace{-0.7cm}
\begin{center}
  \subfloat[\textit{Persistent Change Filter} on Data without Missing Problem]{\includegraphics[width=0.43\textwidth]{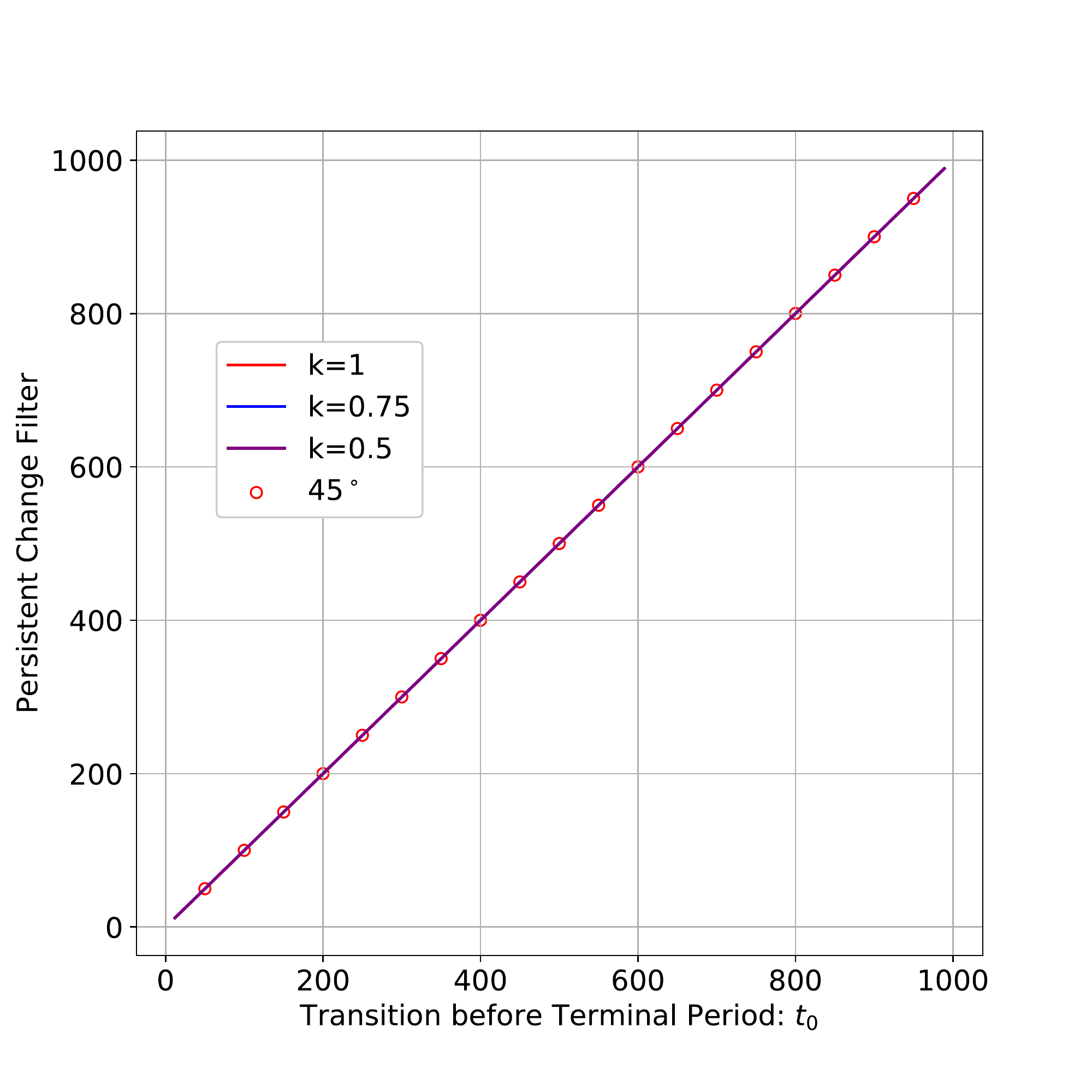}}
  \hfill
  \subfloat[\textit{Persistent Change Filter} on Data with Missing Problems]{\includegraphics[width=0.43\textwidth]{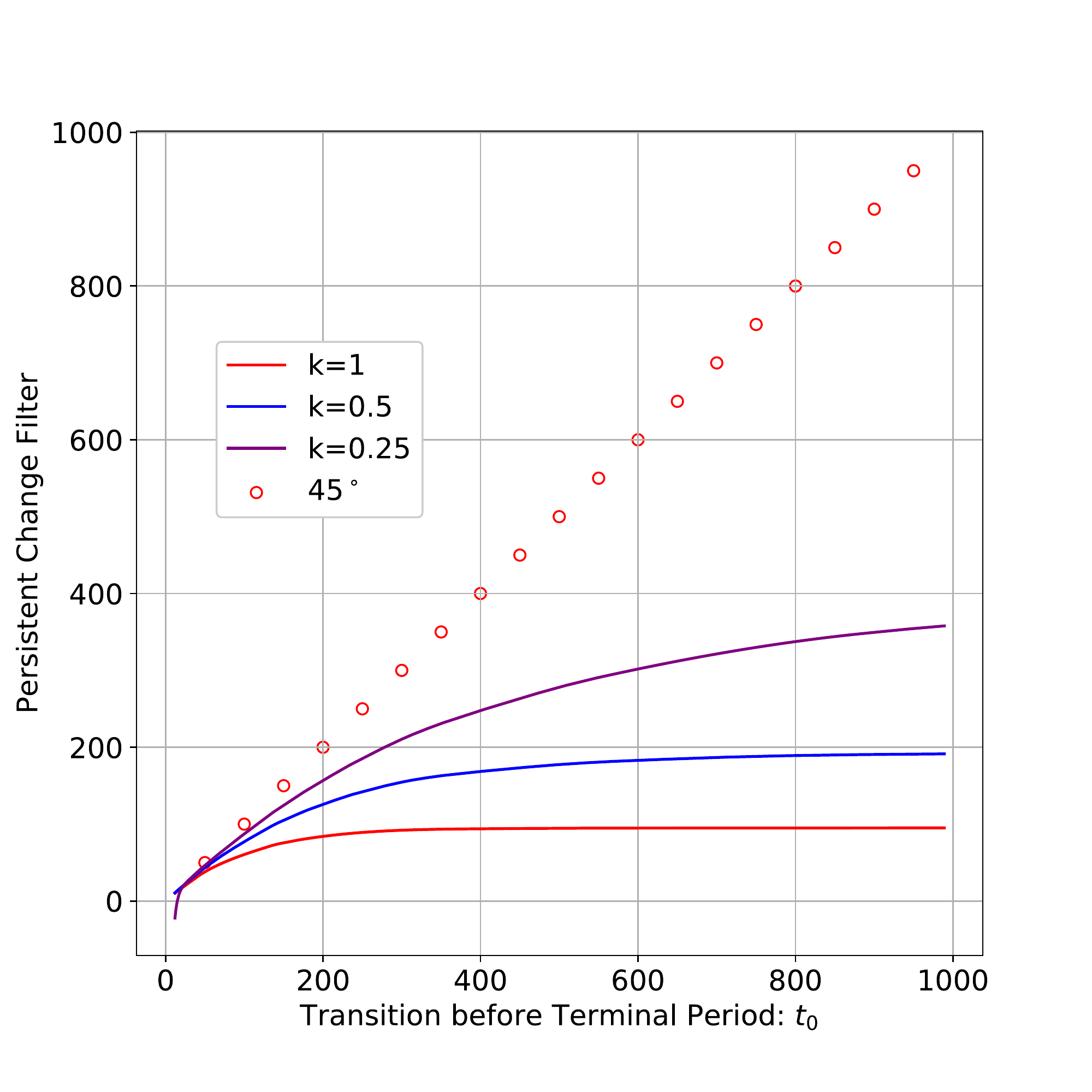}}
  \caption{\textit{Persistent Change Filter} Captures the Duration of a Persistent Change}
    \label{fig:df}
  \end{center}
\end{figure}

\textbf{Example I} For a binary time series $x_t\in \{0,1\},t=1,2...T = 1000$. $x_t$ begins with 0 for some periods, and persistently switch to 1 for the last $t_0$ periods: $x_t = \mathbbm{1}[t > T-t_0]$. We plot the \textit{persistent change filter} on the entire time series $\mathcal{D}(\{x_\tau\}_{\tau = 1}^T)$ against the number of periods $t_0$ between the transition and the terminal period in the left panel of Figure \ref{fig:df}. We find that no matter what smoothing variable $k$ we choose, the \textit{persistent change filter} fits perfectly with $t_0$ on the 45$^{\circ}$ line. This implies that the \textit{persistent change filter} captures the duration of a persistent change in such a binary time series $0,...,0, 1, 1,...,1$.\par
\textbf{Example II} In real world, such perfect binary data may not exist. Consider time series $x_t$, where everything is the same as {Example I}, except that a random 5\% of $x_t$ are replaced with $x_t = 0.9$ when $t > T-t_0$ to mimic an abnormal data or data missing problem. Then we plot the \textit{persistent change filter} $\mathcal{D}(\{x_\tau\}_{\tau = 1}^T)$ against $t_0$ for different choices of $k$ in the right panel of Figure \ref{fig:df}. For the the \textit{persistent change filter} without any smoothing (i.e. $k=1$), the plot deviates from the 45$^{\circ}$ line significantly. When $k$ decreases, the plot gets closer to the 45$^{\circ}$ line. So an optimal choice of $k$ would help capture the duration of a persistent change in such a time series with potential data issues. To note, $k$ would be left as an unknown parameter that our model would learn from the data. \par
We formalize the examples above with the following proposition:
\begin{prop} \label{prop:orig}
When $k\rightarrow 1$, $p_T$ uniformly converges to $\mathcal{D}^{(0)}$ defined as below:
$$\mathcal{D}^{(0)}(x_T,x_{T-1}...x_1)=\sum_{i=1}^{T}\prod_{j=1}^{i}x_{T-j+1}.$$
\end{prop}
What does $\mathcal{D}^{(0)}$ means for the time series? Consider the binary time series $x_t\in \{0,1\},t=1,2...T$, $\mathcal{D}^{(0)}$ is the lasting time periods for recent $x_i$'s to be 1. In other words,
$$\mathcal{D}^{(0)}(x)=m\ \Longleftrightarrow\ x_T=x_{T-1}...=x_{T-m+1}=1,x_{T-m}=0$$
Thus $\mathcal{D}^{(0)}$ gives information on the lasting time since the most recent period when the binary time series persistently switch to 1. For continuous $x_t$ in $[0,1]$, $\mathcal{D}^{(0)}$ is differentiable at any point, which means it could be part of a trainable model. However, $\mathcal{D}^{(0)}$ only captures the jumps from low values to high values, so we extend $\mathcal{D}^{(0)}$ to $\mathcal{D}$ so that it can capture both persistent jumps and drops in the time series\footnote{Some might find the idea of the persistent change filter related to the structural break literature in time series analysis \cite{casini2018structural}. However, the structural break detection methods are not applicable for feature constructions, since they mostly focus on detecting break points with formal statistical tests, while we hope to get an explicit function transform as part of the interpretable model.}. This is the $q_t$ term in the formulation. As is discussed above, we also add the trainable smoothing parameter $k$ to adjust for data missing or abnormal data problems. For a more elaborate discussion and the origins of the \textit{persistent change filter}, please refer to Supplement \ref{app:defn}.

\section{Application}\label{sec:application}
\subsection{Employment Status Prediction with Interpretable Neural Networks}\label{sec:employ}
In this section, we apply the interpretable neural network model for panel data to predict individual's monthly employment status with administrative data. The data come from a four-million population city in China\footnote{For confidentiality, we would not release any city-specific information including the city name in this paper.}, and includes basic demographic information (age, family relations, gender, education, etc.), as well as individual level monthly payments to six different kinds of social insurances and the Housing Provident Fund (HPF)\footnote{In China, there are five major types of social insurances: endowment insurance, basic medical insurance, unemployment insurance, employment injury insurance, and maternity insurance. For urban residents, basic medical insurance consists of ``basic medical insurance for working urban residents'' and ``basic medical insurance for non-working urban residents''. Together with the Housing Provident Fund (HPF), which we take as one type of insurance from now on, our administrative data include detailed individual level payments to seven types of insurances in total. To note, those who pay for employment injury insurance or  unemployment insurance are not necessarily employed or unemployed, and vice versa.}. With all these features, together with employment/unemployment labels on part of the sample (about 400,000 individuals are labelled every month), we construct an interpretable neural network model to predict employment status of all the individuals in the population each month. With the prediction results, we can calculate unemployment rate on the whole population as an important economic indicator for policy makers. This application is novel and important given the unreliability of China's official unemployment statistics \cite{feng2017long}.\par
For individual $i$ in calendar month $t$, her employment status is denoted as $y_{it}\in \{0,1\}$, where 1 and 0 correspond to being employed and unemployed respectively. Her payment amounts for different insurances, among other individual features are denoted as $x_{ijt}$ where $j=1,2...m$ correspond to different features. To predict $y_{it}$, we stack features of the individual in the past $s$ period as model inputs, denote as $\Tilde{\mathbf{x}}_{it}=(x_{ij\tau})$ where $j=1,2...m, \tau = t, t-1, ..., t-s+1$.

Following Section \ref{sec:math}, we write the model as
$P(y_{it}=1|\Tilde{\mathbf{x}}_{it})=g(\Tilde{\mathbf{x}}_{it},\theta)$ where $g=g^{(4)} \circ g^{(3)} \circ g^{(2)} \circ g^{(1)}$. The original input variables of $g$ is a $ms\times 1$ vector:
$$\tilde{\mathbf{x}}_{i t}=\left(x_{i j \tau}\right) \text { where } j=1,2 \ldots m, \tau=t, t-1, \ldots, t-s+1.$$
Here we use monthly payment data of various social insurances and the HPF from July 2016 to December 2017 to construct $\Tilde{\mathbf{x}}_{it}$, so $m = 7$ is the number of insurances to pay and $s = 6$ denotes the lagged periods we consider in model inputs.
With the administrative data in Section \ref{sec:employ}, . To construct a balanced data sample, we randomly select 20,000 employed observations and 20,000 unemployed observations. Both positive and negative samples are evenly divided into two parts to get the training set and the test set.
\subsection{Model Results}\label{sec:results}
To evaluate our model, we will compare it with other interpretable models (logistic regressions) as well as more complicated models (random forests or neural networks). These models require handcrafted features as model inputs. Based on the descriptive statistics and domain knowledge (see the details in Supplement \ref{sec:desc}) on this problem, we construct the following features:
\begin{enumerate}[leftmargin=*]
\item ``Insurance count'' (IC). The number of types of insurances the individual pays in period $t$.
\item ``Naive persistent change'' (NPC). For each kind of insurance, we construct the ``naive persistent change'' $\mathcal{D}^{(0)}(x_t,x_{t-1},...,x_{t-s+1})$ following the definition in Proposition \ref{prop:orig}, where $x_{\tau}$ is a binary variable on whether the individual does not pay\footnote{As is discussed in Proposition \ref{prop:orig}, $\mathcal{D}^{(0)}$ only works for increasing time series ..., 0, 0,..., 0, 1,..., 1. So here we define $x_{\tau}$ as whether the individual does NOT pay, rather than pay  a specific insurance in period $\tau$. } for a specific insurance in period $\tau$. Different from the \textit{persistent change filter}, these ``naive'' measures do not incorporate any trainable parameters, and come directly from feature engineering.
\item ``Payment change'' (PC). We construct another ``naive persistent change'' variable taking the same expression based on the following binary time series $x_\tau$: whether the number of types of insurances the individual has paid is larger than 2. The feature with threshold 2 is constructed with insights from the descriptive statistics (see Supplement \ref{sec:desc}).
\end{enumerate}
With these features in hand, we compare the performance of the following 11 models:
\begin{enumerate}[leftmargin=*]
\item The interpretable neural network (IntNN) model (1) we propose in this paper. No handcrafted feature is needed in this model.
\item Conventional interpretable model: logistic regression (Logistic) with (2) 7 ``naive persistent changes'', ``payment change'' and insurance count; (3) 7 ``naive persistent changes''; (4) ``payment change'' and insurance count; (5) ``payment change''; (6) insurance count.
\item Conventional uninterpretable model: random forests (RF) and neural networks (NN) with (7) 7 ``naive persistent changes'', ``payment change'' and insurance count; (8) 7 ``naive persistent changes''; (9) ``payment change'' and insurance count; (10) ``payment change''; (11) insurance count. 
\end{enumerate}
The predictive accuracy of the best-performing models under each category is reported in Table \ref{table:results1}.\par
\begin{wraptable}{r}{8cm}
\caption{\small{Model Comparisons}}
\label{table:results1}
  \centering
\begin{tabular}{llll}
\toprule
Index & Model            & Inputs                           & Test Accuracy    \\ \hline
1     & IntNN & -                                & \textbf{0.94525} \\ \hline
2 & \multirow{5}{*}{Logistic} & NPC, IC, PC & 0.9284         \\
3     &                  & NPC                  & 0.87035          \\
4     &                  & PC, IC & 0.81905          \\
5     &                  & PC                   & 0.81205          \\ 
6     &                  & IC                  & 0.84435          \\ \hline
7 & \multirow{5}{*}{RF \& NN}    & NPC, IC, PC & \textbf{0.929} \\
8     &                  & NPC                  & 0.929            \\
9     &                  & PC, IC & 0.90605          \\
10    &                  & PC                   & 0.84435          \\
11    &                  & IC                  & 0.86395          \\
\bottomrule
\end{tabular}
\end{wraptable}
Comparing the interpretable neural network model with logistic regression models, we find our model enjoys higher accuracy no matter how we feature engineer the logistic inputs. After  hyperparameter tuning, random forest or conventional neural network models are more accurate than logistic regressions, but are still less accurate than our model, which might be a bit surprising to some readers. This comparison is actually in line with the finding of \cite{rieger2019interp} which argues that the increase in interpretability may allow models to achieve higher accuracy in the test set than unrestricted models.


\subsection{Model Interpretation}\label{sec:interp}
The interpretable neural network models deliver a clear mechanism behind the model outcome. Now we look into the model estimation results for each component of  $g=g^{(4)} \circ g^{(3)} \circ g^{(2)} \circ g^{(1)}$, and interpret the model we obtain.
\begin{enumerate}[leftmargin=*]
\item Decision-tree-like splitting $g^{(1)}$: For $\forall j, \tau$, 
${x}_{i j\tau}^{(1)}=\mbox{Sigmoid}(c{x}_{i j\tau}+d)$.\\
The estimates are $c=2.60,d=-41.46$. As is discussed in Section \ref{sec:math}, with such $c$ and $d$, $g^{(1)}$ would transform small payment values to $0$, while keep large payment values close to 1. So, it approximates an indicator function of whether the individual pays each type of insurances. 
\item Dimension reduction $g^{(2)}$: For $\forall \tau$, ${x}_{i \tau}^{(2)} = \mbox{Sigmoid}(w^T({x}_{i j\tau}^{(1)})_{j = 1}^m+b)$.\\
The estimates of the m-dimensional vectors $w = (-1.73, 1.84, -4, 0.64, 0.62, -0.66, 4.19)^T$, and each element corresponds to endowment insurance, urban working medical insurance, unemployment insurance, employment injury insurance, maternity insurance, urban non-working medical insurance, and the Housing Provident Fund (HPF) respectively. The intercept $b = 3.67$.\par
In this layer, we obtain a 1-dimension variable ${x}_{i \tau}^{(2)}$ from the linear combination of the 7-dimensional payment records after the decision-tree-like splitting. The output is positively correlated with payments of urban working medical insurance, employment injury insurance, maternity insurance, and the Housing Provident Fund (HPF), while negatively correlated with payments of endowment insurance, unemployment insurance, urban non-working medical insurance.\footnote{We also compare $w$ with parameters from logistic regression models, and find they share qualitatively the same interpretation while our model obtains higher accuracy. See Supplement \ref{sec:comp2} for more details.}
\item \textit{Persistent Change Filter} $g^{(3)}$: $\tilde{{x}}_{i t}^{(3)} =\mathcal{D}(({x}_{i \tau}^{(2)})_{\tau = t}^{t-s+1}, k)$.\\
The optimal smoothing parameter learnt from the original data is $k=0.999999$. As a \textit{persistent change filter} on ${x}_{i \tau}^{(2)}$, a large positive $\tilde{{x}}_{i t}^{(3)}$ corresponds to a persistent jump of ${x}_{i \tau}^{(2)}$, while a a large negative value corresponds to a persistent drop of ${x}_{i \tau}^{(2)}$.
\item Logistic regression $g^{(4)}$: $P(y_{it}=1) =\mbox{Sigmoid}(u\tilde{{x}}_{i t}^{(3)}+v)$\\
The estimates are $u=1.06,v=-2.18$. This is a differentiable version of linear transform with coefficient close to 1, which is similar to identity.
\end{enumerate}

To summarize, our interpretable neural network model predicts individual's employment status simply with the \textit{persistent change filter} of a composite variable from a linear combination of whether an individual pay each type of insurance. The weights of the linear combination imply the relative importance of each insurance when predicting the employment status. From the model, we learn that when an individual that used to consistently pay urban working medical insurance, employment injury insurance, maternity insurance, or the Housing Provident Fund (HPF), but suddenly drop out from those insurance programs, there is a higher probability for her to become unemployed. Similarly, for individuals that are beginning to get enrolled in endowment insurance, unemployment insurance, urban non-working medical insurance, there is a higher probability for her to get unemployed. These model interpretations would be helpful for policymakers who hope to understand the accurate model predictions and design policies based on machine learning models.

\subsection{Robustness of the Model}
As is discussed in \cite{murdoch2019definitions}, an interpretable model should be robust to data missing or abnormal data problem. To mimic such scenarios, we randomly set 10\% of all the payment records to 0, and train the same 11 models as in Section \ref{sec:results}. The accuracy comparison results still hold. Also, the parameter estimates deliver qualitatively equivalent interpretations. The value of parameter $k$ in the \textit{persistent change filter} becomes smaller to offset data missing. Please refer to Supplement \ref{sec:missing} for more details.
\section{Conclusion}
In this paper, we propose a class of interpretable neural network models that can achieve both high prediction accuracy and interpretability. We first propose a general formulation of such neural networks, which begins with interpretable functional forms encoded in the first several layers, followed by the final layer with a regularized number of interpretable features as input. The model satisfies all the three desiderata for interpretable machine learning models \cite{murdoch2019definitions}. Researchers can design different forms of interpretable functions based on the nature of their tasks. In this paper, we incorporate a class of interpretable functions named \textit{persistent change filters} as part of the neural network to study panel data in economics.\\

As an application, we use the model to predict individual's monthly employment status using high-dimensional administrative data in China. We achieve a high accuracy as 94.5\% on the test set, which is comparable to the best-performing conventional machine learning methods. Furthermore, we interpret the model to understand the mechanisms behind. We find it accurately predicts individual's employment status simply with the \textit{persistent change filter} of a composite variable from a linear combination of whether an individual pay each type of insurance. The model deliver robust interpretations subject to potential data issues, and would be helpful for researchers and policymakers to understand useful information contained in the big administrative data.\par

This paper contributes to the machine learning toolbox of economists by introducing a modified version of neural networks that is both accurate and easy to interpret. The model can encode different interpretable functional forms that researchers may propose based on the nature of their own research tasks. For example, researchers who study default risks may propose specific functional forms that are interpretable on the loan data. With the massive use of administrative and proprietary big data in economics and policy research, lots of research could be done with this new tool.\par


\bibliographystyle{plain}

\begin{thebibliography}{10}
	
	\bibitem{abadie2019using}
	Alberto Abadie.
	\newblock Using synthetic controls: Feasibility, data requirements, and
	methodological aspects.
	\newblock {\em Journal of Economic Literature}, 2019.
	
	\bibitem{angrist2008mostly}
	Joshua~D Angrist and J{\"o}rn-Steffen Pischke.
	\newblock {\em Mostly harmless econometrics: An empiricist's companion}.
	\newblock Princeton university press, 2008.
	
	\bibitem{athey2018impact}
	Susan Athey.
	\newblock The impact of machine learning on economics.
	\newblock In {\em The economics of artificial intelligence: An agenda}, pages
	507--547. University of Chicago Press, 2018.
	
	\bibitem{athey2018matrix}
	Susan Athey, Mohsen Bayati, Nikolay Doudchenko, Guido Imbens, and Khashayar
	Khosravi.
	\newblock Matrix completion methods for causal panel data models.
	\newblock Technical report, National Bureau of Economic Research, 2018.
	
	\bibitem{casini2018structural}
	Alessandro Casini and Pierre Perron.
	\newblock Structural breaks in time series.
	\newblock {\em arXiv preprint arXiv:1805.03807}, 2018.
	
	\bibitem{chernozhukov2018double}
	Victor Chernozhukov, Denis Chetverikov, Mert Demirer, Esther Duflo, Christian
	Hansen, Whitney Newey, and James Robins.
	\newblock Double/debiased machine learning for treatment and structural
	parameters, 2018.
	
	\bibitem{chetty2014land}
	Raj Chetty, Nathaniel Hendren, Patrick Kline, and Emmanuel Saez.
	\newblock Where is the land of opportunity? the geography of intergenerational
	mobility in the united states.
	\newblock {\em The Quarterly Journal of Economics}, 129(4):1553--1623, 2014.
	
	\bibitem{doshi2017towards}
	Finale Doshi-Velez and Been Kim.
	\newblock Towards a rigorous science of interpretable machine learning.
	\newblock {\em arXiv preprint arXiv:1702.08608}, 2017.
	
	\bibitem{einav2014economics}
	Liran Einav and Jonathan Levin.
	\newblock Economics in the age of big data.
	\newblock {\em Science}, 346(6210), 2014.
	
	\bibitem{feng2017long}
	Shuaizhang Feng, Yingyao Hu, and Robert Moffitt.
	\newblock Long run trends in unemployment and labor force participation in
	urban {C}hina.
	\newblock {\em Journal of Comparative Economics}, 45(2):304--324, 2017.
	
	\bibitem{giles2005china}
	John Giles, Park Albert, and Juwei Zhang.
	\newblock What is {C}hina's true unemployment rate?
	\newblock {\em China Economic Review}, 16(2):149--170, 2005.
	
	\bibitem{kingma2014adam}
	Diederik~P Kingma and Jimmy Ba.
	\newblock Adam: A method for stochastic optimization.
	\newblock {\em arXiv preprint arXiv:1412.6980}, 2014.
	
	\bibitem{kleinberg2018human}
	Jon Kleinberg, Himabindu Lakkaraju, Jure Leskovec, Jens Ludwig, and Sendhil
	Mullainathan.
	\newblock Human decisions and machine predictions.
	\newblock {\em The Quarterly Journal of Economics}, 133(1):237--293, 2018.
	
	\bibitem{lei2017interpretable}
	Tao Lei.
	\newblock {\em Interpretable neural models for natural language processing}.
	\newblock PhD thesis, Massachusetts Institute of Technology, 2017.
	
	\bibitem{mian2015house}
	Atif Mian and Amir Sufi.
	\newblock {\em House of debt: How they (and you) caused the Great Recession,
		and how we can prevent it from happening again}.
	\newblock University of Chicago Press, 2015.
	
	\bibitem{mullainathan2017machine}
	Sendhil Mullainathan and Jann Spiess.
	\newblock Machine learning: an applied econometric approach.
	\newblock {\em Journal of Economic Perspectives}, 31(2):87--106, 2017.
	
	\bibitem{murdoch2019definitions}
	W~James Murdoch, Chandan Singh, Karl Kumbier, Reza Abbasi-Asl, and Bin Yu.
	\newblock Definitions, methods, and applications in interpretable machine
	learning.
	\newblock {\em Proceedings of the National Academy of Sciences},
	116(44):22071--22080, 2019.
	
	\bibitem{rieger2019interp}
	Laura Rieger, Chandan Singh, W~James Murdoch, and Bin Yu.
	\newblock Interpretations are useful: penalizing explanations to align neural
	networks with prior knowledge.
	\newblock {\em arXiv preprint arXiv:1909.13584}, 2019.
	
	\bibitem{saez2016wealth}
	Emmanuel Saez and Gabriel Zucman.
	\newblock Wealth inequality in the united states since 1913: Evidence from
	capitalized income tax data.
	\newblock {\em The Quarterly Journal of Economics}, 131(2):519--578, 2016.
	
	\bibitem{tibshirani1996regression}
	Robert Tibshirani.
	\newblock Regression shrinkage and selection via the lasso.
	\newblock {\em Journal of the Royal Statistical Society: Series B
		(Methodological)}, 58(1):267--288, 1996.
	
	\bibitem{wager2018estimation}
	Stefan Wager and Susan Athey.
	\newblock Estimation and inference of heterogeneous treatment effects using
	random forests.
	\newblock {\em Journal of the American Statistical Association},
	113(523):1228--1242, 2018.
	
	\bibitem{wooldridge2016introductory}
	Jeffrey~M Wooldridge.
	\newblock {\em Introductory econometrics: A modern approach}.
	\newblock Nelson Education, 2016.
	
	\bibitem{yang2018deep}
	Yongxin Yang, Irene~Garcia Morillo, and Timothy~M Hospedales.
	\newblock Deep neural decision trees.
	\newblock {\em arXiv preprint arXiv:1806.06988}, 2018.
	
	\bibitem{yuan2006model}
	Ming Yuan and Yi~Lin.
	\newblock Model selection and estimation in regression with grouped variables.
	\newblock {\em Journal of the Royal Statistical Society: Series B (Statistical
		Methodology)}, 68(1):49--67, 2006.
	
	\bibitem{zhang2018interpretable}
	Quanshi Zhang, Ying Nian~Wu, and Song-Chun Zhu.
	\newblock Interpretable convolutional neural networks.
	\newblock In {\em Proceedings of the IEEE Conference on Computer Vision and
		Pattern Recognition}, pages 8827--8836, 2018.
	
\end{thebibliography}

\newpage
\appendix
\section*{Supplementary Materials}

\section{Descriptive Statistics}\label{sec:desc}
Table \ref{table:desc} presents some descriptive statistics on insurance payment behaviors of both the employed and the unemployed sample.
\begin{table}[h!]
\caption{\small{Descriptive Statistics}}
\label{table:desc}
\centering
\vspace{7pt}
\begin{tabular}{|c|c|c|c|c|}
\toprule
\multicolumn{5}{|c|}{\# of people that pay each insurance or HPF in the most recent period} \\ \hline
            & endowment  & urban working medical           & unemployment  & injury \\ \hline
employed    & 10938      & 13737             & 19750         & 13503  \\ \hline
unemployed  & 15026      & 0                 & 19510         & 0      \\ \hline
            & maternity  & non-working medical & HPF &        \\ \hline
employed    & 14167      & 0                 & 9356          &        \\ \hline
unemployed  & 0          & 42                & 443           &        \\ \hline
\multicolumn{5}{|c|}{\# of types of insurances and HPF paid in the most recent period}       \\ \hline
            & 0          & 1                 & 2             & 3      \\ \hline
employed    & 0          & 5523              & 347           & 632    \\ \hline
unemployed  & 369        & 4582              & 14708         & 341    \\ \hline
            & 4          & 5                 & 6             & 7      \\ \hline
employed    & 140        & 7370              & 5988          & 0      \\ \hline
unemployed  & 0          & 0                 & 0             & 0      \\ 
\bottomrule
\end{tabular}
\end{table}
The upper panel is the number of employed and unemployed individuals that pay each type of the insurances in the most recent month in the sample. From this panel we can see that the unemployed individuals tend to pay endowment insurance, unemployment insurance and urban working medical insurance, while the employed individuals pay all the insurances except for urban non-working medical insurance, since they mostly prefer urban working medical insurance. The lower panel reports the number of types of insurances that paid by each employed and unemployed individual in the most recent month, we find that the employed individuals tend to pay more kinds of insurances. With the elbow method, we find 2 is a good threshold to discriminate the employed and the unemployed if we hope to do some traditional feature engineering, since most of the unemployed sample pay less or equal to 2 types of insurances.

\section{Interpretation Comparison: Our Model vs Logistic Regression}\label{sec:comp2}
We compare our model with the logistic regression model with the ``naive persistent changes'' of seven kinds of insurances as model inputs. The definition of ``naive persistent change'' is in Section \ref{sec:results}. The linear combination weights in $g^{(2)}$ of our interpretable neural network model and the logistic coefficients of each variable are in Table \ref{table:compare}.\par
\begin{table}[h]
\caption{Comparison: Interpretable Neural Networks vs. Conventional Interpretable Models}
\label{table:compare}
\centering
\vspace{7pt}
\begin{tabular}{|c|c|c|c|c|c|c|c|}
\toprule
                 & Endowment & Working medical & Unemploy & Injury & Maternity & Non-work medical & HPF   \\ \hline
IntNN & -1.73     & 1.84            & -4           & 0.64   & 0.62      & -0.66               & 4.19  \\ \hline
Logistic    & 2.02      & -0.61           & 0.32         & -0.52  & -1.16     & 1.53                & -0.12 \\
\bottomrule
\end{tabular}
\end{table}
From Table \ref{table:compare}, we find the interpretations of our model and the logistic regression model are qualitatively the same. As is discussed in the definition of ``naive persistent change'' in Section \ref{sec:results}, larger naive persistent change means the individual switch from paying to not paying a specific type of insurance. Thus negative coefficients in the logistic regression imply a positive relationship between jumps in payments and the probability getting employed, which share the same interpretation as positive coefficients in our model. Here we find the signs of all the coefficients in the interpretable neural networks are exactly opposite to each other, which means they share exactly the same interpretation qualitatively.
\section{Robustness of the Model}\label{sec:missing}
\subsection{Model Accuracy with Data Missing}\label{app:int_missing}
The model we propose is robust to data missing or abnormal data problems. To check the robustness, we randomly set 10\% of all the payment records to 0 as data missing. Then we rerun the same 11 models as in Section \ref{sec:results}. The model accuracy results are in Table \ref{table:results2}.\par
\begin{table}[h!]
\caption{Robust Model Comparisons: Our Model vs. Logistic Regression vs. Random Forest}
\label{table:results2}
\centering
\vspace{7pt}
\begin{tabular}{llll}
\toprule
Index & Model            & Inputs                           & Test Accuracy    \\ \hline
1     & IntNN & -                                & \textbf{0.92855} \\ \hline
2 & \multirow{5}{*}{Logistic} & NPC, IC, PC & 0.91735         \\
3     &                  & NPC                  & 0.85805          \\
4     &                  & PC, IC & 0.79235         \\
5     &                  & PC                   & 0.7781          \\ 
6     &                  & IC                  & 0.84025          \\ \hline
7 & \multirow{5}{*}{RF \& NN}    & NPC, IC, PC & \textbf{0.92715} \\
8     &                  & NPC                  & 0.917            \\
9     &                  & PC, IC & 0.85645          \\
10    &                  & PC                   & 0.84025          \\
11    &                  & IC                  & 0.84025          \\
\bottomrule
\end{tabular}
\end{table}
From Table \ref{table:results2}, we can see that all the results are similar subject to the 10\% data missing. Though data missing makes all the models worse off, our interpretable model remain the best performance among all the competitors. In the next section, we will see the model parameters are also robust, which delivers robust model interpretation for the problem.
\subsection{Model Interpretation with Data Missing}\label{app:int_missing2}
In Section \ref{sec:interp}, we interpret our model clearly through the model structure. Besides transparency of the model structure, robustness is another important feature of model interpretability. In this section, we illustrate the robustness of our model with model parameters we estimate with 10\% missing data in Section \ref{sec:missing}. 
\begin{enumerate}[leftmargin=*]
\item Decision-tree-type splitting $g^{(1)}$: For $\forall j, \tau$, 
${x}_{i j\tau}^{(1)}=\mbox{Sigmoid}(c{x}_{i j\tau}+d)$. 
The estimates are $c=2.76,d=-43.96$, which are both close to those in the baseline model.
\item Dimension reduction $g^{(2)}$: For $\forall \tau$, $
{x}_{i \tau}^{(2)} = \mbox{Sigmoid}(w^T({x}_{i j\tau}^{(1)})_{j = 1}^m+b)$.
The estimates of the m-dimensional vectors $w = (-0.76, 1.57, -2.12, 0.55, 0.5, -0.8, 3.39)^T$. The intercept $b = 2.87$. All the estimates are close to those in the baseline model.
\item \textit{Persistent Change Filter} $g^{(3)}$: 
$\tilde{{x}}_{i t}^{(3)} =\mathcal{D}(({x}_{i \tau}^{(2)})_{\tau = t}^{t-s+1}, k)$.
The estimated smoothing parameter is $k=0.89$, different from the baseline estimates 0.999999. This is totally sensible: due to data missing, we need to impose some smoothing to get a reasonable \textit{persistent change filter} time series that predict the outcomes well. This also confirms the rationale of the \textit{persistent change filter} definition, as is detailed discussed in Supplement \ref{app:defn}.
\item Logistic regression $g^{(4)}$:
$P(y_{it}) =\mbox{Sigmoid}(u\tilde{{x}}_{i t}^{(3)}+v)$.
The estimates are $u=1.82, v=-5.96$, which are both close to those in the baseline model.
\end{enumerate}
To summarize, our model is quite robust subject to data missing problems, and deliver essentially the same interpretations of the model outcomes. 

\section{Motivation and Variants of \textit{Persistent Change Filter}}\label{app:defn}
The \textit{persistent change filter} is a monotone reduction designed for univariate time-series input $x=(x_1,x_2...x_T),\ x_t\in [0,1]$ and mainly captures the persistent jumps or drops in it.
\subsection{Persistent change $\mathcal{D}^{(0)}$ for Binary Inputs}
Consider the binary time series $x_t\in \{0,1\},t=1,2...T$, persistent change is defined as the lasting time periods for recent $x_i$'s to be 1. In other words,
$\mathcal{D}^{(0)}(x)=m\ \Longleftrightarrow\ x_T=x_{T-1}...=x_{T-m+1}=1,x_{T-m}=0$.
This variable gives information on the moment the input turns into 1 and the lasting time.
\subsection{Continuous Persistent change $\mathcal{D}^{(1)}$ for Continuous Inputs}
When we have a continuous input $x_t\in[0,1], t=1, 2..., T$, we find a compatible expression for persistent change which reduces to the original version if we restrict the input to be binary.
$$\mathcal{D}^{(1)}=\sum_{t=1}^{T}\prod_{j=1}^{t}x_{T-j+1}$$
In an iterative way, we can define a series ${p_t}, t = 1,2...,T$ and 
$p_1=x_1,p_{t+1}=x_{t+1}+x_{t+1}p_t$,
then we have the continuous persistent change measure
$\mathcal{D}^{(1)}=p_T$.
To understand the iterative definition, we see when $x_{t+1}$ gets to 0, it will almost clean up the historical accumulation $p_{t}$. When $x_{t+1}$ remains to be 1, it will keep the historical accumulation $p_t$ and $p_{t+1}$ accumulates by adding $x_{t+1}$.
\subsection{Persistent change $\mathcal{D}^{(2)}$: Symmetrical to Jumps and Drops}\label{sec:app_sym}
The measures $\mathcal{D}^{(0)}$ and $\mathcal{D}^{(1)}$ can only capture time series change from small values (like 0) to big values (like 1). To treat small and large values equally, we define a new measure $\mathcal{D}^{(2)}$ on continuous value time series $x_t\in[0,1], t=1, 2..., T$:
$$p_1=x_1,p_{t+1}=x_{t+1}+x_{t+1}p_{t}; \space q_1=1-x_1,q_{t+1}=(1-x_{t+1})+(1-x_{t+1})q_{t}$$
$$\mathcal{D}^{(2)}=p_T-q_T$$
In this definition, both jumps and drops in $x_{t}$ are addressed.
\subsection{Smooth Persistent change $\mathcal{D}^{(3)}$: Adding a trainable smooth variable $k$}
To weaken the cleaning-up effect against abnormal time series change due to data missing or data error problems, we define a smoothing variable $k \in [0,1]$, and define a new measure $\mathcal{D}^{(3)}$ on continuous value time series $x_t\in[0,1], t=1, 2..., T$:
$$p^{s}_1=x_1,p^{s}_{t+1}=x_{t+1}+ kx_{t+1}p_{t} + (1-k)p_{t}$$
$$q^{s}_1=1-x_1,q^{s}_{t+1}=(1-x_{t+1})+k(1-x_{t+1})q^{s}_{t}+(1-k)q^{s}_{t}$$
$$\mathcal{D}^{(3)}=p^{s}_T-q^{s}_T$$
Compared to $\mathcal{D}^{(2)}$, the multiplication term $x_{t+1}p_{t}$ is replaced by $kx_{t+1}p_{t}+(1-k)p_{t}$. In other words, larger $k$ means larger cleaning-up effect, while the accumulation effect is unaffected.  When $k = 1$ we have $\mathcal{D}^{(3)} = \mathcal{D}^{(2)}$. We leave $k$ trainable to let the data finds the best value. Finally, we have the notion of \textit{persistent change filter} $\mathcal{D}=\mathcal{D}^{(3)}$.

\subsection{Why Do We Call It \textit{Persistent Change Filter}?}
Consider four pairs of inputs:
$$x_1^{(1)}=(0,0,0,1,1,1,1,1,1,1,1,1,1,1,1),x_1^{(2)}=(1,1,1,0,0,0,0,0,0,0,0,0,0,0,0);$$
$$x_2^{(1)}=(0,0,0,1,1,1,1,1,1,1,\mathbf{0},1,1,1,1),x_2^{(2)}=(1,1,1,0,0,0,0,0,0,0,\mathbf{1},0,0,0,0);$$
$$x_3^{(1)}=(0.1,0.1,0.1,0.9,0.9,0.9,...,0.9,0.9), x_3^{(2)}=(0.9,0.9,0.9,0.1,0.1,0.1,...,0.1,0.1);$$
$$x_4^{(1)}=(0.1,0.1,0.1,0.9,0.9,0.9,0.9,0.9,0.9,0.9,\mathbf{0.1},0.9,0.9,0.9,0.9),$$
$$x_4^{(2)}=(0.9,0.9,0.9,0.1,0.1,0.1,0.1,0.1,0.1,0.1,\mathbf{0.9},0.1,0.1,0.1,0.1);$$
and the persistent change measures we defined above are in Table \ref{table:distance}.
\begin{table}[h!]
\caption{Persistent Change Measures of Example Time Series}
\label{table:distance}
\centering
\vspace{7pt}
\begin{tabular}{|l|l|l|l|l||l|l|l|l|l|}
\hline
    & $\mathcal{D}^{(0)}$ & $\mathcal{D}^{(1)}$ & $\mathcal{D}^{(2)}$ & $\mathcal{D}^{(3)},k=0.25$&     & $\mathcal{D}^{(0)}$ & $\mathcal{D}^{(1)}$ & $\mathcal{D}^{(2)}$ & $\mathcal{D}^{(3)},k=0.25$ \\ \hline
$x_1^{(1)}$ & 12           & 12           & 12           & 11.90            &
$x_1^{(2)}$ & 0           & 0           & -12          & -11.90          \\ \hline
$x_2^{(1)}$ &  4          & 4          & 4          & 8.81           &
$x_2^{(2)}$ &  0        & 0           & -4         & -8.81          \\ \hline
$x_3^{(1)}$ & --           & 6.49           & 6.37           & 9.06             &
$x_3^{(2)}$ & --           & 0.11           & -6.37          & -9.06           \\ \hline
$x_4^{(1)}$ & --           & 3.47          & 3.36          & 6.90           &
$x_4^{(2)}$ & --          & 0.111           & -3.36          & -6.90          \\ \hline
\end{tabular}

\end{table}

From Table \ref{table:distance}, we have the following intuitive findings as below:
\begin{enumerate}[leftmargin=*]
\item As is shown in the first two rows, $\mathcal{D}^{(1)}$ is equivalent to $\mathcal{D}^{(0)}$ for binary inputs, while it can also handle continuous inputs as the last two rows.
\item $\mathcal{D}^{(2)}$ captures the persistent change idea, but is vulnerable to abnormal data points, as is in the second and forth rows.
\item $\mathcal{D}^{(3)}$ can still capture the persistent change idea even upon abnormal data points. Here the smoothing parameter $k$ is selected arbitrarily, while it could be trained to an optimal value when we hope to use $\mathcal{D}^{(3)}$ as input features in regression models.
\end{enumerate}

\end{document}